\documentstyle[12pt,titlepage]{article}
\input epsfig.sty 

\renewcommand{\theequation}{\thesection.\arabic{equation}}

\font\grande=cmr10 scaled \magstep4
\font\medio=cmr10 scaled \magstep2
\outer\def\beginsection#1\par{\medbreak\bigskip
      \message{#1}\leftline{\bf#1}\nobreak\medskip\vskip-\parskip
      \noindent}

\def\laq{\raise 0.4ex\hbox{$<$}\kern -0.8em\lower 0.62 ex\hbox{$\sim$}}
\def\gaq{\raise 0.4ex\hbox{$>$}\kern -0.7em\lower 0.62 ex\hbox{$\sim$}}
\def\a{\alpha}

\begin{document}
\bibliographystyle {unsrt}

\titlepage
\vspace{15mm}
\begin{center}
{\grande Regular Cosmological Examples}\\
\vspace{5mm}
{\grande of Tree-Level Dilaton-Driven Models}\\
\vspace{10mm}
\centerline{Massimo Giovannini\footnote{e-mail: M.Giovannini@damtp.cam.ac.uk}}
\bigskip
\centerline{{\it DAMTP, 21 Silver Street,Cambridge, CB3 9EW,
United Kingdom}}
\vskip 2  cm

{\medio  Abstract} \\
\end{center}
\noindent
 We construct  analytic solutions of  the 
 low energy (i.e. tree-level) string cosmological effective
 action. We work with  the ``minimal'' field content
(i.e. graviton and dilaton) in the absence of any dilaton potential
 and of any antisymmetric tensor field. 
Provided the metric is sufficiently inhomogeneous we obtain
 solutions whose  curvature invariants are bounded
 and  everywhere defined in time and space. The dilaton coupling and
 its associated energy density are regular and homogeneous. 
 A phase of growing (and non-singular) dilaton coupling is
 compatible with the regularity of the curvature invariants without
 the addition of higher curvature (or higher genus) corrections to the
 tree-level effective action. We discuss the symmetries and the
 physical properties of the obtained solutions.
\newpage 
\renewcommand{\theequation}{1.\arabic{equation}}
\setcounter{equation}{0}
\section{Introduction and motivations}

A quite intriguing aspect of cosmological models based on the low
energy string theory effective action is the occurrence of a phase of
growing dilaton coupling that might trigger  a number of interesting 
implications which were specifically addressed in the framework of the
pre-big-bang inflationary models \cite{ven1}.
One of the distinctive aspects of this class of models is the
amplification, via gravitational instability, 
of the vacuum inhomogeneities associated both with the metric
 and with the Abelian gauge fields 
\cite{m2}. If this picture is correct, we should live, at later times
in an Universe filled with stochastic backgrounds of gravitational
waves and of of Abelian gauge fields which, thanks to the magnetic
flux conservation in a hot plasma, might also have interesting effects. 

A lot of efforts have certainly been devoted to the analysis of
the cosmological solutions of the low energy string theory effective
action \cite{LTCS} in order to understand if the physical properties
of those solutions were at all relevant for the  explanation of some general
features of our Universe. In all these studies one of the key
assumptions was the complete homogeneity
 of the classical evolution of
the geometry. It has been recently realized that 
it would be quite interesting to
relax this very stringent assumption and, therefore, 
inhomogeneous string cosmological models were discussed
from different points of view \cite{barrow,ven2} but always in the
framework of the low energy string theory effective action.
There are of course different motivations for these investigations.
In general, it is always quite dangerous to make
strict assumptions about the form of the cosmological solutions. We
 deal here with solutions of a completely non-linear system of
evolution equations and we cannot exclude, in principle, physical
situations were the Universe was very inhomogeneous at early times
evolving, ultimately, towards a homogeneous (and possibly isotropic) 
epoch at later times. In order to address these  questions one
should start, from the very beginning, with a completely inhomogeneous
metric. 
Completely inhomogeneous configurations  are also required in
the analysis of the initial conditions of cosmological (inflationary)
models. To relax the homogeneity assumption turned out to be relevant
for understanding if (and when) inflation could take place 
starting from more generic initial conditions. 
An example in this direction are the investigations
 concerning  the role played by initial (spherical) inhomogeneities
in  chaotic inflationary models  \cite{piran}.
With the same motivations (but in the context of string inspired
cosmological models) quasi-homogenous metrics were used 
for the discussion of  the initial value problem 
\cite{ven2} in the pre-big-bang scenario \cite{ven1,gasven}. 
In this last class of models,
the discussion of the initial conditions in a
completely homogeneous metric can lead to quite different results 
\cite{turner}. We argue, therefore, that there is an urgent need
of a better understanding of completely inhomogeneous configurations
of the massless string modes. 

We cannot a priori exclude that the features of
inhomogeneous string cosmological models might be significantly
 different from the ones of their completely homogeneous counterpart.

An open question of (homogeneous) string cosmological models  
 is the occurrence of
curvature singularities
\cite{gasven} which turned out to be closely connected with the
problem of a graceful exit from a phase of growing dilaton coupling
\cite{brusven}, towards a phase of declining dilaton coupling. In 
 a completely homogeneous metrics quite general (numerical)
 studies suggested that a phase of growing coupling is compatible
 with the regularity of the curvature invariants (in the String frame)
 provided higher order (curvature) corrections to the low energy
 (tree-level) effective action are included in the game
 \cite{brusven}. This idea received recently new attention \cite{mm}. 
 Moreover, the idea of a graceful exit transition from the initial 
 phase of growing coupling to the subsequent standard
 radiation-dominated  evolution has been the subject of many
 investigations \cite{prr}.

In this paper we will not address the problem of the graceful exit but
we will  be concerned with the regularity of the curvature
invariants.

In the case of a completely inhomogeneous metric is it still
mandatory to hit a singularity after a phase of growing coupling?
Is it possible to regularize the curvature invariants if the Universe
was sufficiently inhomogeneous at early times?

These are  the main question we will try to address and, in this sense, our
approach is qualitatively different from the ones followed in Ref. 
\cite{mm,prr} (where homogeneity was assumed) 
and could be related to some pioneering 
attempts discussed in Ref. \cite{GMV}.
Regular (inhomogeneous) solutions of the low energy string
theory effective action were constructed \cite{GMV} by ``boosting away'' the
curvature singularities through the $O(d,d)$
symmetries which were extensively studied in the past \cite{odd,meiss} and
which represent a  natural generalization of the scale factor duality
\cite{ven1,sfd}.  The key observation was that starting with a generic singular
cosmology in $1+1$ dimensions or with a $1+1$ dimensional black hole,
the singularity gets ``boosted away'' by $O(2,2)$ transformation involving
a third (originally flat) dimension. Moreover, by  considering the
inhomogeneous cosmological model of Nappi and Witten \cite{NW} it was
also shown that the singularities of the model can be ``boosted away'' by
a $O(3,3)$ transformations involving a fifth dimension.
The main conclusion of Ref. \cite{GMV} was that, provided the coupling
oscillates and provided an antisymmetric tensor background is
included, the curvature invariants are regular. Therefore, there are
no doubts that it is certainly physically relevant to relax the
homogeneity assumption usually made in string inspired cosmological
models. Following this logical line and applying the Occam razor
we want to study inhomogeneous string cosmological models
 with the minimal field content
(i.e. the gravi-dilaton action). In other words, 
we want to understand if it is possible to have
simultaneously satisfied these requirements: growing coupling,
 regularity of the curvature invariants, absence of antisymmetric
tensor field, absence of any dilaton potential.
In this sense the aim  of our investigation is different from (but
complementary to) the one of Ref. \cite{GMV}.

To tackle this problem many different strategies can be certainly
employed. The most general one would consist in trying to solve the
low energy beta functions for a generic metric $g_{\mu\nu}(\vec{x},t)$
depending simultaneously upon  {\em space and time}. A simplification
adopted in previous studies \cite{ven2,barrow} was also to stick to a
{\em diagonal} inhomogeneous metric. 
Also the form of the inhomogeneous metric is, in a way, ``minimal''.
In fact the simplest non-trivial case of fully inhomogeneous 
metrics is the one
where the inhomogeneities  are  distributed along
a line. In this case the whole spatial dependence is reduced to only
one coordinate (which we took, in our examples, to run along the the $x$ axis).
Needless to say that this is not  the most general situation. At the
same time there are no reasons to forbid this choice.
A posteriori we finally found useful to impose an extra symmetry on 
 the metric which will turn out to possess, in our discussion, two
 killing vector fields which are hypersurface orthogonal and
 orthogonal to each other. This form of the
 metric seems to be widely used in the study of inhomogeneous
 cosmological models in general relativity \cite{kra}. 
 It is interesting to notice (as we will explicitly
 show) that this ansatz for the
 metric is compatible with a completely {\em homogenous} dilaton
 coupling which turns out to be, ultimately, only dependent upon time.

We solved 
exactly the low energy beta functions in the String
frame by using all the assumptions listed above. We computed
the curvature invariants associated with the obtained solutions and we
found a class of regular (and parity-invariant) solutions 
  where the curvature invariants are
all bounded. These solutions describe the physical situation where a
growing dilaton coupling smoothly evolves between two asymptotically
constant regimes. The  physical parameter describing the solutions is
essentially the maximal energy density of the dilaton background in
string units.

From a technical point
of view, we have to mention that the most difficult part of our present
exercise is to solve the off-diagonal component of the beta functions
which do not contain second order derivatives with respect to time and
which is then  a constraint mixing first order {\em time} derivatives
with first order {\em spatial} derivatives.
Therefore, the first step will be the  solution of 
 the constraint. If the obtained solution will be compatible with the
dilaton equation and with the remaining components of the beta
functions, then, we will have a particular solution whose singularity
properties can be investigated by the direct calculation of the
curvature invariants.

Before closing this introduction we want to recall that the work
presented in this paper is certainly overlapping with previous work
performed in the context of general relativity. In particular in the
limit of constant dilaton field our solutions match the well known
 inhomogeneous vacuum solutions of general relativity. The
simplest inhomogeneous vacuum models are those with two space-like
commuting killing vectors, known as orthogonally transitive $G_2$
cosmologies \cite{v1,v2}. Moreover, useful background material for the
present investigation can be also found in \cite{verd} where soliton
solutions were discussed in space-times with two space-like Killing
fields.

The plan of our paper is then the following. In Section 2 we will
derive the explicit form of the evolution equations of the dilaton and
of the metric in the String frame picture.
 In Section 3 we will briefly derive  the dilaton vacuum
solutions in the String frame and
we will generalise them to
the case of homogeneous dilaton background with inhomogeneous metric.
In Section 4 we will focus our attention on the case of growing
 dilaton solutions and we will provide two physical examples whose
main physical aspect is the parity invariance of the background geometry.
Motivated the results of Section 4 we generalized our examples to a
class of parity even solutions (Section 5) and we computed the
curvature invariants in this situation. We found that they are regular
as in the case of the examples of Section 4.
Section 6 contains our concluding remarks.
Because of the excessive
length  of the formulas contained in this paper we made the choice of
reporting few (indeed relevant) technical  results in the Appendix which then
collects  useful calculations for the interested reader.

\renewcommand{\theequation}{2.\arabic{equation}}
\setcounter{equation}{0}
\section{Basic Equations} 

The low energy string theory effective action can be written in the
String frame and in the absence of antisymmetric tensor as \cite{LTCS}:
\begin{equation}
S= - \frac{1}{\lambda^2_{s}}
\int d^4 x \sqrt{-g} e^{-\phi} \biggl[ R + g^{\alpha\beta}
\partial_{\alpha}\phi\partial_{\beta} \phi \biggr],
\label{action}
\end{equation}
where $\lambda_{s}$ is the string scale. We assume that the dilaton potential
is completely negligible. We shall consider only the particular
case of critical superstring theory with vanishing cosmological
constant and six (frozen) internal dimensions.
The evolution equations for the massless modes
 can be easily derived by varying the action
with respect to the dilaton field and the metric:
\begin{eqnarray}
& & R - g^{\alpha\beta}\partial_{\alpha}\phi\partial_{\beta}\phi + 2
\Box \phi=0,
\label{dilaton}\\
& & R_{\mu\nu} - \frac{1}{2} g_{\mu\nu} R + \frac{1}{2}g_{\mu\nu}
g^{\alpha\beta} \partial_{\alpha}\phi\partial_{\beta}\phi + 
\nabla_{\mu}\nabla_{\nu}\phi - g_{\mu\nu} \Box\phi =0
\label{equation}
\end{eqnarray}
($\nabla_{\alpha}$ is the covariant derivative with respect to
the metric $g_{\mu\nu}$; 
$\nabla_{\alpha}\nabla_{\beta}\phi =
\partial_{\alpha}\partial_{\beta}\phi - \Gamma_{\alpha\beta}^{\sigma}
 \partial_{\sigma}\phi$; $\Gamma_{\alpha\beta}^{\sigma}$ are the
Christoffel symbols).
Using Eq. (\ref{dilaton})  into Eq. (\ref{equation}) we find the
familiar form of the tree-level beta functions
\begin{eqnarray}
& & R - g^{\alpha\beta}\partial_{\alpha}\phi\partial_{\beta}\phi + 2
\Box \phi=0,
\label{beta1}\\
& & R_{\mu}^{\nu} + \nabla_{\mu}\nabla^{\nu} \phi=0.
\label{beta2}
\end{eqnarray}
We now specialize to the case of a diagonal metric which admits two
commuting space-like killing vector fields, both of which are
hypersurface orthogonal \cite{kra}:
\begin{equation}
~g_{00}= A(x,t),~~~ g_{x x} = -A(x, t), ~~~g_{yy} = - B(x, t) C(x, t), ~~~
g_{zz}= - \frac{B(x,t)}{C(x,t)}.
\label{metric}
\end{equation}
The corresponding line element is then
\begin{equation}
ds^2 = A(x,t)[ dt^2 - dx^2] - B(x,t)\biggl[ C(x,t) dy^2 +
\frac{dz^2}{C(x,t)}\biggr].
\label{line}
\end{equation}
From Eqs. (\ref{metric}) and (\ref{line})
 the Christoffel symbols, the Riemann tensors and  the
Weyl tensors can be easily computed. In Appendix A we report the
Christoffel symbols, the Ricci tensors and the curvature scalar which
are required in order to express directly  
Eqs. (\ref{beta1}) and (\ref{beta2}) in the metric (\ref{metric}). 
In the metric (\ref{metric}), Eqs (\ref{beta1}) and (\ref{beta2}) will
produce a non-linear system of (partial) differential equations
which we are going to solve. First of all  suppose that the
space-time dependence of the 
metric functions can be factorised:
\begin{equation}
A(x,t) = a(t)~\alpha(x),~~~~B(x,t) = b(t)~\beta(x),~~~~
C(x,t) = c(t)~\kappa(x),
\label{ansatz}
\end{equation}
(notice that the latin letters will denote the temporal part whereas the Greek
ones will denote the spatial part of the metric functions).
This decomposition has the great advantage, once inserted in
Eqs. (\ref{beta1}) and (\ref{beta2}), of transforming the original
system into  a system of
non-linear {\em ordinary} differential equations in the two variables
$x$ and $t$.
By inserting Eq (\ref{ansatz}) into the components of the Ricci
tensors and into the curvature scalar given in Appendix A
 (see Eqs. (\ref{rxx})-(\ref{r})) we can 
write down explicitly Eq. (\ref{beta1}) and all the components of 
Eq. (\ref{beta2}) in the metric given in Eq. (\ref{metric}). 
The $(xx)$, $(yy)$, $(zz)$ and $(00)$ components of Eq. (\ref{beta2}) will be 
respectively:
\begin{eqnarray}
&&\frac{\ddot{a}}{a} + \frac{\dot{a}}{a} \frac{\dot{b}}{b}
-\left(\frac{\dot{a}}{a}\right)^2 - \frac{\dot{a}}{a} \dot{\phi} = 2
\frac{\beta''}{\beta}+\left(\frac{\alpha'}{\alpha}\right)' +
\left(\frac{\kappa'}{\kappa}\right)^2 - \left(\frac{\beta'}{\beta}\right)^2 -
\frac{\alpha'}{\alpha}~\frac{\beta'}{\beta} ~~~~~~~~~~(xx),
\label{xx}\\
&&\frac{\ddot{c}}{c} + \frac{\ddot{b}}{b} + \frac{\dot{b}}{b}
\frac{\dot{c}}{c} - \left(\frac{\dot{c}}{c}\right)^2 - \left(\frac{\dot{b}}{b} 
+ \frac{\dot{c}}{c}\right)\dot{\phi} = \frac{\kappa''}{\kappa}
+\frac{\beta''}{\beta} 
- \left(\frac{\kappa'}{\kappa}\right)^2 
+ \frac{\beta'}{\beta}~\frac{\kappa'}{\kappa}~~~~~~~~~~~~~~(yy),
\label{yy}\\
&&\frac{\ddot{b}}{b} + (\frac{\dot{c}}{c})^2
-\frac{\dot{b}}{b}~\frac{\dot{c}}{c} - \frac{\ddot{c}}{c} -
\left(\frac{\dot{b}}{b} -\frac{\dot{c}}{c}\right)\dot{\phi}= 
\frac{\beta''}{\beta} - \frac{\kappa''}{\kappa} + (\frac{\kappa'}{\kappa})^2
-\frac{\beta'}{\beta}\frac{\kappa'}{\kappa}~~~~~~~~~~~~~~~~~(zz),
\label{zz}\\
&&2 \ddot{\phi} -  \frac{\dot{a}}{a} \dot{\phi}  +
\left(\frac{\dot{a}}{a}\right)^2 +\frac{\dot{a}}{a}~\frac{\dot{b}}{b}+
\left(\frac{\dot{b}}{b}\right)^2 - \left(\frac{\dot{c}}{c}\right)^2 -
 \frac{\ddot{a}}{a} 
- 2 \frac{\ddot{b}}{b} = 
- \frac{\alpha'}{\alpha}~\frac{\beta'}{\beta} 
- \left(\frac{\alpha'}{\alpha}\right)' ~~~~(00), 
\label{00}
\end{eqnarray}
whereas for Eq. (\ref{beta1}) and for the mixed
 component (i.e. $(x0)$) of Eq. (\ref{beta2}) we will have respectively:
\begin{eqnarray}
&&2 \ddot{\phi} + 2 \frac{\dot{b}}{b}\dot{\phi} - {\dot{\phi}}^2 
+ \frac{1}{2} \left(\frac{\dot{b}}{b}\right)^2 
-  \frac{1}{2} \left(\frac{\dot{c}}{c}\right)^2
- \frac{d}{dt}\left(\frac{\dot{a}}{a}\right) - 2 \frac{\ddot{b}}{b} = 
\frac{1}{2}(\frac{\beta'}{\beta})^2 
- \frac{1}{2}\left(\frac{\kappa'}{\kappa}\right)^2
\nonumber\\
&&- \left(\frac{\alpha'}{\alpha}\right)' - 2
\frac{\beta''}{\beta}~~~~~~~~~~~~~~~~~~~~~~~~~~~~~
~~~~~~~~~~~~~~~~~~~~~~~~~~~~~~~~~~~~~~~~~~(\phi)
\label{phi}\\
&&\frac{\dot{b}}{b} ~\frac{\alpha'}{\alpha} + \frac{\dot{a}}{a}~
\frac{\beta'}{\beta} - \frac{\beta'}{\beta}~ \frac{\dot{b}}{b} -
\frac{\dot{c}}{c}~ \frac{\kappa'}{\kappa}-
\frac{\alpha'}{\alpha}\dot{\phi} = 0
\label{x0}~~~~~~~~~~~~~~~~~~~~~~~~~~~~~~~~~~~~~~~~(x0)
\end{eqnarray}
(in these equations and in all the paper we adopted the compact
 notation : $'= \frac{\partial}{\partial x}$ and $\cdot= 
\frac{\partial}{\partial t}$).
In the system of Eqs. (\ref{xx})--(\ref{phi}) at the left hand side 
 of each equation we only have {\em time}-dependent quantities whereas
  at the right hand side we only have {\em space}-dependent quantities.
The exception is represented by Eq. (\ref{x0}) which is 
 {\em not} a dynamical equation but  a
constraint mixing spatial and temporal derivatives. 
The coupled system of non-linear differential equations is written in
the case where the dilaton coupling is completely {\em homogeneous},
namely $\phi' =0,~~~~\phi(x,t) \equiv \phi(t)$.
The solution of the system of coupled, non-linear differential
equations derived in the previous equations will be the main goal of
the next Section.

\renewcommand{\theequation}{3.\arabic{equation}}
\setcounter{equation}{0}
\section{Homogeneous dilaton solutions} 

The first step in the derivation of our solutions with homogeneous
dilaton will be to consider the constant dilaton case where $\dot{\phi}(t)=0$.
In order to find a general expression able to solve the constraint, 
we sum and subtract
Eqs. (\ref{yy}) and (\ref{zz}), obtaining, respectively,
\begin{equation}
\frac{\ddot{b}}{b} = \frac{\beta''}{\beta},~~~~
\Biggl[\frac{d}{d t}\left(\frac{\dot{c}}{c}\right) 
+ \frac{\dot{b}}{b}\frac{ \dot{c}}{c}\Biggr] = \Biggl[\frac{d}{dx} \left(
\frac{\kappa'}{\kappa}\right) +
\frac{\beta'}{\beta}\frac{\kappa'}{\kappa} \Biggr].
\label{condition}
\end{equation}
A particular solution of
Eq. (\ref{condition}) which might solve  the constraint is
\begin{equation}
b(t)= \cosh{\mu t},~~~~~\beta(x) = \sinh{\mu x}.
\label{ans}
\end{equation}
It is true that Eq. (\ref{condition}) can be solved also for another
choices of trial functions: not all the possible choices are allowed
by the off-diagonal constraint given in Eq. (\ref{x0}), which can now
be re-written as
\begin{equation}
\Biggl[\mu \tanh{\mu t} \left(\frac{\alpha'}{\alpha}\right) +
\left(\frac{\dot{a}}{a}\right)\frac{\mu}{\tanh{\mu x}} - \mu^2 \frac{\tanh{\mu 
t}}{\tanh{\mu x}} - \left(\frac{\dot{c}}{c}\right) 
\left(\frac{\kappa'}{\kappa}\right)\Biggr]=0.
\label{constr}
\end{equation}
Provided
\begin{equation}
\frac{\dot{a}}{a} = c_{1}\tanh{\mu t},~{\rm and}~~~\frac{\dot{c}}{c}= c_{2}
\tanh{\mu t},
\label{ansatz1}
\end{equation}
the time dependence  appearing in Eq.
(\ref{constr}) factorizes
leading, ultimately, to a first-order (ordinary) differential relation
involving only functions depending upon the spatial
coordinates. Therefore direct (trivial) integration of 
Eqs. (\ref{condition}) and (\ref{constr}), together with the conditions
(\ref{ans}) and (\ref{ansatz1}), gives us the following first
integrals  
\begin{eqnarray}
&&\left(\frac{\kappa'}{\kappa}\right) = \frac{c_{2} +
c_{3}}{2}\biggl(\tanh{\frac{\mu x}{2}}\biggr)^{-1} + \frac{c_{2}-c_{3}}{2}
\biggl(\tanh{\frac{\mu x}{2}}\biggr),
\label{kappa}\\
&&\left(\frac{\alpha'}{\alpha}\right) = \Biggl[
\biggl(\frac{c^2_{2}+c_{2} c_{3} - \mu
 c_{1} + \mu^2}{2 \mu}\biggr)\biggl(\tanh{\frac{\mu x}{2}}\biggr)^{-1} 
+ \biggl(\frac{c^2_{2} - c_{2} c_{3} - \mu
c_{1} +\mu^2 }{2\mu}\biggr) \tanh{\frac{\mu x}{2}}\Biggr],
\label{alpha}
\end{eqnarray}
where $c_{3}$ is now an integration constant and 
$c_{1}$ and $c_{2}$ are nothing but arbitrary constants which have to be
fixed by requiring the compatibility of Eqs. (\ref{ans}) and
(\ref{ansatz1}) with the complete system given by by Eqs. 
(\ref{xx})--(\ref{phi}). 
Let us therefore find the compatibility relations which have to be
satisfied by $c_{1}$, $c_{2}$ and $c_{3}$. 
By direct substitution of Eqs. (\ref{ans}), (\ref{ansatz1}), (\ref{kappa}) and 
 (\ref{alpha}) into Eqs. (\ref{xx}) and (\ref{00}) we get, respectively
\begin{equation}
c^2_{3} = 3 \mu^2 + c^2_{2} - 2 \mu c_{1},~~~c^2_{2} =2 \mu c_{1} + \mu^2.
\label{c3}
\end{equation}
Surprisingly enough the same two conditions also solve
Eq. (\ref{phi}). Moreover Eqs. (\ref{yy}), (\ref{zz}) and (\ref{x0})
are automatically satisfied for every $c_{1}$, $c_{2}$ and $c_{3}$
not leading (as expected) to any further relation among the
constants appearing in our trial solution.
The present solutions are the String frame version of
analogous inhomogeneous vacuum solutions previously discussed in the
context of general relativity \cite{kra,v1}. 

After having solved the case where the dilaton was constant we can now
discuss the case where the dilaton is time dependent.
 The ansatz given in Eqs. (\ref{ans}) and (\ref{ansatz1}) can be
slightly modified in order to
match the time dependence of the dilaton field, which now appears, in
the equations of motion (\ref{00})--(\ref{phi}) 
 with its first and second derivatives.
The idea is then to assume that in the presence of the dilaton field
the solutions of our system of equations can be written as
\begin{equation}
\left(\frac{\dot{a}}{a}\right) = c_{1} \tanh{\mu t} +
\Gamma(t),~~~\left(\frac{\dot{b}}{b}\right)= \mu \tanh{\mu t} + \Delta(t), ~~~
\left(\frac{\dot{c}}{c}\right) = c_{2} \tanh{\mu t} + \Sigma(t)~~,
\label{trial}
\end{equation}
where, at the moment $\Gamma(t)$, $\Delta(t)$ and $\Sigma(t)$, are
three unknown functions which should be hopefully determined by
solving explicitely the evolution equations of the metric and of the
dilaton field.
By comparing this ansatz with one made  in the previous Section (see
Eq. (\ref{ansatz1})) it is evident which kind of logic we
followed since for $\Gamma=\Delta=\Sigma=0$ we exactly recover our
Eq. (\ref{ansatz1}) . Notice that $\Gamma$, $\Delta$ and $\Sigma$ are only
dependent upon time since we want to accommodate a {\em homogenous}
dilaton .

With the ansatz (\ref{trial}) we then repeat step by step the
procedure outlined in Section 1. First of all, we insert
Eq. (\ref{trial}) into Eq. (\ref{x0}) and we get the condition which
needs to be satisfied in order to solve the constraint:
\begin{eqnarray}
\Biggl[\left(\frac{\alpha'}{\alpha}\right) + \frac{ \mu - c_{1}}{\tanh{\mu x}} \Biggr]= \frac{c_{2}}{\mu}~
\left(\frac{\kappa'}{\kappa}\right) 
- \frac{1}{\mu~\tanh{\mu t}}\Biggl\{ \frac{\mu~(\Gamma
- \Delta)}{\tanh{\mu x}} -
\left(\frac{\kappa'}{\kappa}\right)~ \Sigma - 
\left(\frac{\alpha'}{\alpha}\right)~ \dot{\phi}\Biggr\}.
\label{first}
\end{eqnarray}
By  summing  and subtracting Eqs. (\ref{yy}) and  (\ref{zz}), we obtain 
\begin{equation}
\frac{\ddot{b}}{b} -\frac{\dot{b}}{b}\dot{\phi} =
\frac{\beta''}{\beta},~~~~\Biggl[\frac{d}{d t}\left(\frac{\dot{c}}{c}\right) 
- \frac{\dot{b}}{b}\frac{ \dot{c}}{c} +\frac{\dot{c}}{c} \dot{\phi}\Biggr] 
= \Biggl[\frac{d}{dx} \left(
\frac{\kappa'}{\kappa}\right) +
\frac{\beta'}{\beta}\frac{\kappa'}{\kappa}\Biggr].
\label{I}
\end{equation}
Inserting Eq. (\ref{trial}) into Eqs. (\ref{I}),  we
find, respectively:
\begin{eqnarray}
&&\dot{\Delta} + [ \Delta - \dot{\phi}]~\Delta + \mu~\tanh{\mu t}~ [ 2~
\Delta - \dot{\phi}] =0,
\label{second}\\
&&\Biggl[\frac{d}{d x}\left( \frac{\kappa'}{\kappa}\right) +
\frac{\beta'}{\beta} \frac{\kappa'}{\kappa} -
c_{2}\mu\Biggr] = \dot{\Sigma} + \tanh{\mu t}~ [ \mu~ \Sigma + c_{2}~ ( \Delta
- \dot{\phi})] + \Sigma~(\Delta - \dot{\phi}).
\label{third}
\end{eqnarray}
Eqs. (\ref{first}), (\ref{second}) and (\ref{third}) can be
consistently solved if $\dot{\phi}(t) = \Delta(t) =
\Gamma(t),~~~\Sigma=0$ leading to the solution
\begin{eqnarray}
\left(\frac{k'}{k}\right) &=& \Biggl[\frac{c_{3} +c_{2}}{2}\left(
\tanh{\frac{\mu x}{2}}\right)^{-1} 
+ \frac{c_{2} - c_{3}}{2}\tanh{\frac{\mu x}{2}}\Biggr],
\nonumber\\
\left(\frac{\alpha'}{\alpha}\right)&=&\Biggl[ \frac{c_{2}^2 
+ c_{2}c_{3} - \mu c_{1} +
\mu^2}{2 \mu} \left( \tanh{\frac{\mu x}{2}}\right)^{-1}  +
  \frac{c_{2}^2 + c_{2}c_{3} - \mu c_{1} + \mu^2}{2
\mu}\tanh{\frac{\mu x}{2}}\Biggr],
\nonumber\\
\Gamma(t) &=& \frac{c_{4}}{\cosh{\mu t}},
\label{system2}
\end{eqnarray}
where $c_{3}$ and $c_{4}$ are integration constants.
Substituting Eqs. (\ref{trial}) and (\ref{system2}) 
into Eq. (\ref{xx}) and   Eq. (\ref{00}) we
obtain, respectively, the following compatibility conditions
\begin{equation}
c^2_{3} = 3 \mu^2 + c^2_{2} - 2 \mu c_{1} \geq 0,~~~
c^2_{4} = c^2_{2} - \mu^2 - 2 \mu c_{1} \geq 0.
\label{rel2}
\end{equation}
Finally going into Eq. (\ref{phi}) we deduce that
\begin{equation}
2c_{4}^2 - \mu^2 + 6 \mu c_{1} = 3 c_{2}^2 - c_{3}^2,
\label{rel3}
\end{equation}
which turns out to be noting but a trivial linear combination of the
two conditions given in Eq. (\ref{rel2}).
As expected, Eqs. (\ref{yy}) and (\ref{zz}) are then satisfied without
imposing any further condition on $c_{1}$, $c_{2}$,  $c_{3}$ and
$c_{4}$. 

We are now ready to write down the explicit form of our solutions, namely:
\begin{eqnarray}
A(x,t) &=& e^{c_{4} gd(\mu t)}~ \Biggl[\cosh{\mu
t}\Biggr]^{\frac{c_1}{\mu}}
~\Biggl[\sinh{\left(\frac{\mu x}{2}\right)}
\Biggr]^{\frac{c_2^2 + c_2 c_3 - \mu c_1 + \mu^2}{\mu^2}}~
\Biggl[\cosh{\left(\frac{\mu x}{2}\right)}
\Biggr]^{ \frac{c_2^2 - c_2 c_3 - \mu c_1
+ \mu^2}{\mu^2} },
\nonumber\\
B(x,t) &=&  e^{c_{4} gd(\mu t)}~\cosh{\mu t}~ \sinh{\mu x},
\nonumber\\
C(x,t) &=&\Biggl[\cosh{\mu t}\Biggr]^{\frac{c_2}{\mu}}~ 
\Biggl[\sinh{\left(\frac{\mu x}{2}\right)}\Biggr]^{\frac{c_{2} + c_{3}}{\mu}}~ 
\Biggl[\cosh{\left(\frac{\mu x}{2}\right)}\Biggr]^{\frac{c_{2} - c_{3}}{\mu}},
\nonumber\\
\phi(t) &=& \frac{c_{4}}{\mu}~gd(\mu t) - c_{5},
\label{thesolution}
\end{eqnarray}
where $gd(\mu t)= \arctan{[\sinh{\mu t}]}$ 
is the ``Gudermannian'' \cite{ryz} or hyperbolic amplitude. Notice
that the limit $c_{4}\rightarrow 0$ is slightly ill defined since, from
Eq. (\ref{rel2}), $c_2$ might take imaginary values and the
inequalites are, in this limit, strict equalities. Therefore, the
general expression of the solution (\ref{thesolution}) also changes,
in this limit.

Since the hyperbolic amplitude always grows, the dilaton
will either increase or decrease depending upon the sign of $c_{4}$.
Therefore, the physical meaning of $c_{5}$ is connected with the
initially small (perturbative) value of the coupling constant, whereas
$c_4$ is connected with its ``kinetic'' energy. The 
physical situation we are describing with this solutions is really a
regime of very small dilaton coupling and in this sense we will
fine-tune $c_{5}$ to a quite small value. In this way higher loop
(genus) corrections will be automatically subleading. This procedure
is very reminiscent of what happens in the pre-big-bang scenario 
\cite{ven1} where a (quite long) small  coupling regime emerges
naturally from the tree-level (Kasner-like)  solutions. In the pre-big-bang
case, the Kasner-like solutions have two ``branches'' separated by a
(curvature) singularity, and then, for each growing coupling solution
there is also a decreasing coupling solution. 
We notice that, also in this sense, our solutions have close
analogies to the pre-big-bang case since, depending upon the value of
$c_{4}$ there are two branches: one of increasing coupling
($c_{4}>0$) and the other of decreasing coupling ($c_{4}<0$). 
In the pre-big-bang case the singularity  cannot be removed 
(at tree-level).  We will show, in the next
two sections, that in our solutions the two branches are not
analytically connected but, at the same time, the curvature
invariants, the dilaton kinetic energy are always well defined and
regular for any
$x,t$ at least for a family of solutions contained in
Eq. (\ref{thesolution}) (see Sections 4 and 
 5 for a discussion of this point). 

\renewcommand{\theequation}{4.\arabic{equation}}
\setcounter{equation}{0}
\section{Two physical examples} 

From Eq. (\ref{thesolution}) each particular solution of the system of
Eqs. (\ref{beta1}) and (\ref{beta2}) can be specified by fixing
$c_{1}$ and $c_{2}$ and by computing $c_{3}$ and $c_{4}$ from
Eqs. (\ref{rel2}). We tried different possible sets
of parameters and we found that not {\em all} the choices
of $c_{1}$ and $c_{2}$ will necessarily lead to the regularity of the
curvature invariants. The class of
regular solutions seems to be connected to the behaviour of the metric
tensor under (discrete) parity transformations (see the following
Section). At the same
time we point out that {\em there are} choices of $c_{1}$ and $c_{2}$
which lead to completely regular curvature invariants.
In this Section we discuss then particular examples of solutions 
contained in Eq. (\ref{thesolution}) that share this quite interesting
property. In order to do that we will be inspired by some physical
requirement, namely we would like to have regularity of the curvature
invariants and, simultaneously, growing dilaton coupling which
corresponds to $\dot{\phi} > 0$.

For the set of parameters given by
\begin{equation}
c_{1} = 6 \mu,~~~c_{2}= 5 \mu,~~~c_{3} = - 4 \mu, ~~~c_{4}= 2\sqrt{3} \mu,
\label{ex1a}
\end{equation}
Eqs.  (\ref{rel2}) are clearly solved.
The explicit form of the solution of Eq. (\ref{thesolution}) can then
be written, using Eqs. (\ref{ex1a}), as:
\begin{eqnarray}
&&A(x,t) = e^{2~\sqrt{3}~ gd(\mu t)}~ \Biggl[\cosh{\mu t}\Biggr]^{6}~
\Biggl[\cosh{\left(\frac{\mu x}{2}\right)}\Biggr]^{40},~~~~
B(x,t) =  e^{2~\sqrt{3}~ gd(\mu t)}~\cosh{\mu t}~ \sinh{\mu x},
\nonumber\\
&&C(x,t) =\Biggl[\cosh{\mu t}\Biggr]^{5}~ 
\Biggl[\sinh{\left(\frac{\mu x}{2}\right)}\Biggr]~ 
\Biggl[\cosh{\left(\frac{\mu x}{2}\right)}\Biggr]^{9},
~~~~\phi(t) = 2\sqrt{3}~gd(\mu t) - c_{5}.
\label{thesolution1}
\end{eqnarray}
The dilaton is an increasing function
whose maximal value is  given by $\phi(+\infty)= \sqrt{3}\pi - c_{5}$
(recall that $\lim_{t\rightarrow - \infty} gd(\mu t) = - (3/2) \pi$
and that $\lim_{t\rightarrow +\infty}  gd(\mu t) =  (\pi/2)$).
Now, our discussion is based on the (tree-level) action given by
Eq. (\ref{action}) which is valid provided 
$g(\phi) \equiv \exp{[\phi/2]} = \exp{[\frac{c_{4}}{2} gd(\mu t) -
c_{5}]}\ll 1$.
If this is not the case, higher genus corrections should be
included. Consequently, for the compatibility of the solutions
discussed in the present Section with the effective description
adopted in Section 2 we have to require that the dilaton starts its
time evolution deep in its perturbative regime and this can be
achieved by fixing $c_{5}$. In the example given by the solution
(\ref{thesolution1}) we have that $g(+\infty) = \exp{( \sqrt{3} \pi
-c_{5})}\ll 1$, which implies that $c_{5} \gg \sqrt{3} \pi$. Therefore
choosing, for instance $c_{5} \sim 10 \sqrt{3} \pi$ we get that
$g(-\infty) = \exp{(- 13\sqrt{3} \pi)}$ and 
$g(+\infty) = \exp{(-9\sqrt{3}\pi)}$.

Consider now a further example, namely
\begin{equation}
c_{1} = 5 \mu,~~~c_{2}= 4\mu,~c_{3}= - 3\mu,~~{\rm},~~~c_{4}=\sqrt{5}\mu.
\label{ex2a}
\end{equation}
Using Eqs. (\ref{ex2a}) into Eq. (\ref{thesolution})
we get
\begin{eqnarray}
&&A(x,t) = e^{\sqrt{5}~ gd(\mu t)}~ \Biggl[ \cosh{\mu
t}\Biggr]^{5}
\Biggl[\cosh{\left(\frac{\mu x}{2}\right)}\Biggr]^{24},
~~~~B(x,t) =  e^{\sqrt{5}~ gd(\mu t)}~\cosh{\mu t}~ \sinh{\mu x},
\nonumber\\
&&C(x,t) =\Biggl[\cosh{\mu t}\Biggr]^{4}~ 
\Biggl[\sinh{\left(\frac{\mu x}{2}\right)}\Biggr]~ 
\Biggl[\cosh{\left(\frac{\mu x}{2}\right)}\Biggr]^{7},
~~~~\phi(t) = \sqrt{5}~gd(\mu t) - c_{5}.
\label{thesolution2}
\end{eqnarray}

The calculation of the curvature invariants for the solutions 
(\ref{thesolution1}) and (\ref{thesolution2}) can be now performed.
In order to summarise our results we can say that each of the
curvature invariants  
 can be written (in a generalised notation) as:
\begin{equation}
{\cal I }= \mu^4~ f(x,t) [\cosh{\mu t}]^{-a}~ 
\Biggl[\cosh{\frac{\mu x}{2}}\Biggr]^{-b}
\label{inv}
\end{equation}
(${\cal I}$ labels a generic curvature invariant; 
$f(x,t)$ is a regular expression containing combinations of
 hyperbolic functions and it changes for each specific invariant 
[see Appendix B];
 $a$ and $b$ also change depending
upon the invariant under consideration, [see also Table 1]). 

\begin{figure}
\begin{center}
\begin{tabular}{|c|c|}
      \hline
      \hbox{\epsfxsize = 7.5 cm  \epsffile{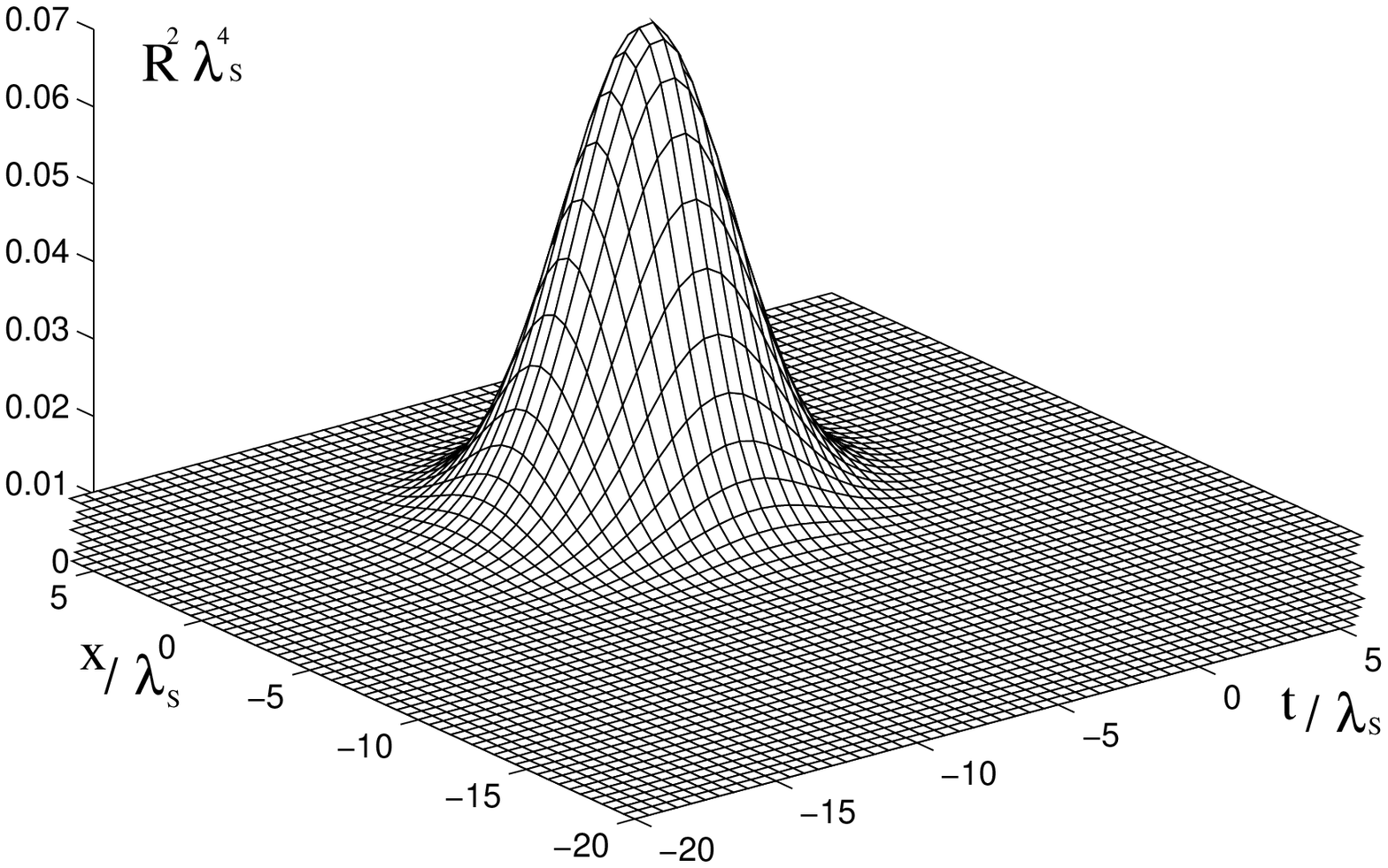}} &
      \hbox{\epsfxsize = 7.5 cm  \epsffile{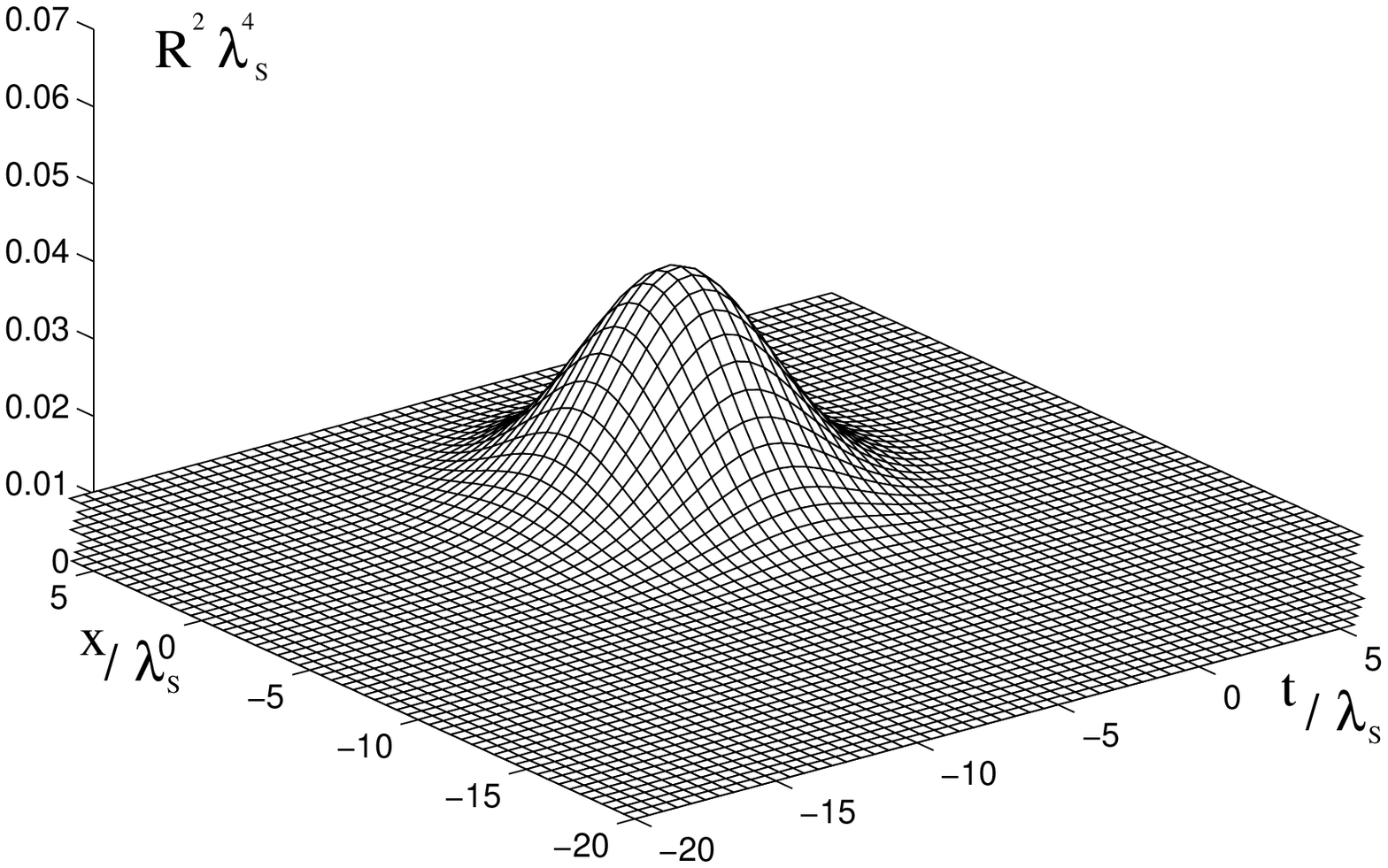}}\\
      \hline
\end{tabular}
\end{center}
\caption{We plot the $R^2$ invariants corresponding
 to the solution (\ref{thesolution1}) (left) and to the 
solution (\ref{thesolution2})(right). The analytic form of these
 invariants is given  in Appendix B (Eq. (\ref{scalar})) by choosing
 $\a = 2\sqrt{3}$ (for the plot at the left) and $\a=  \sqrt{5}$ (for
 the plot at the right). Here we also took $L= 10~\lambda_{s}$}
\label{rsquare}
\end{figure}

\begin{figure}
\begin{center}
\begin{tabular}{|c|c|}
      \hline
      \hbox{\epsfxsize = 7.5 cm  \epsffile{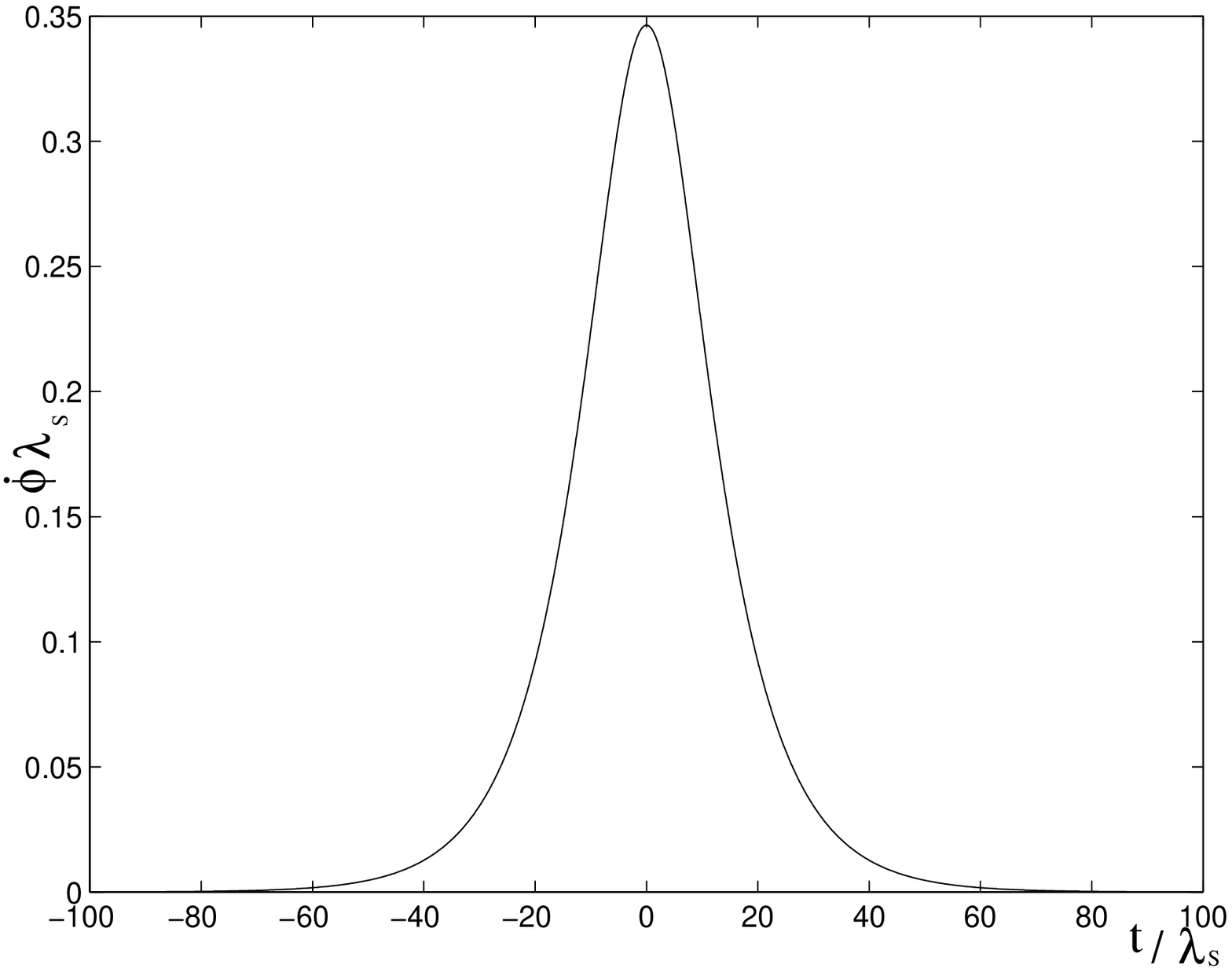}}&
      \hbox{\epsfxsize = 7.5 cm  \epsffile{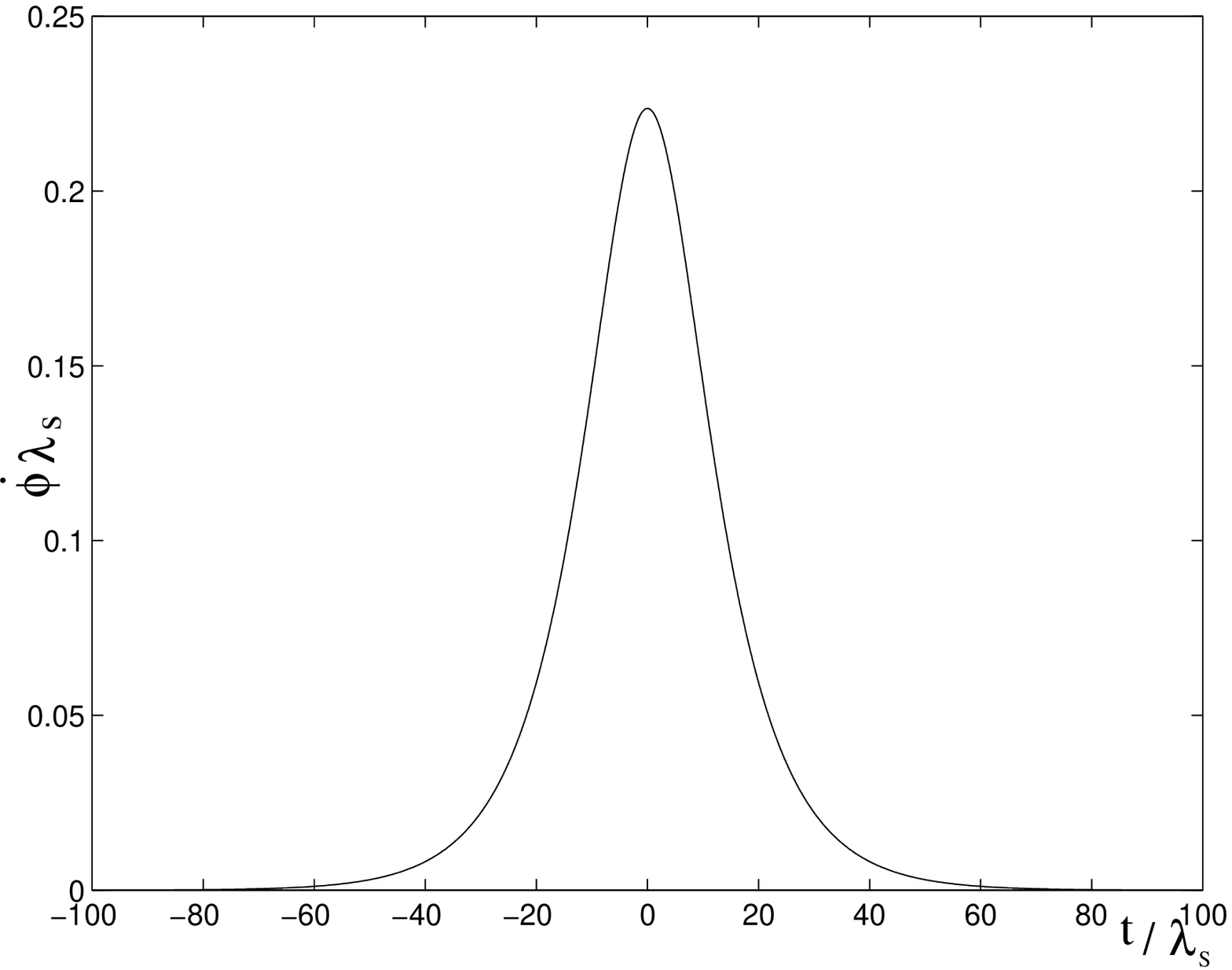}}\\
      \hline
\end{tabular}
\end{center}
\caption{We plot the dilaton kinetic term for the two exact
solutions discussed in this Section. The left figure corresponds
to the case given in Eq. (\ref{thesolution1}), whereas the right one
corresponds to Eq. (\ref{thesolution2}). We notice the absence of
 any singularity.}
\label{dilatonfig}
\end{figure}

\begin{figure}
\begin{center}
\begin{tabular}{|c|c|}
      \hline
      \hbox{\epsfxsize = 7.5 cm  \epsffile{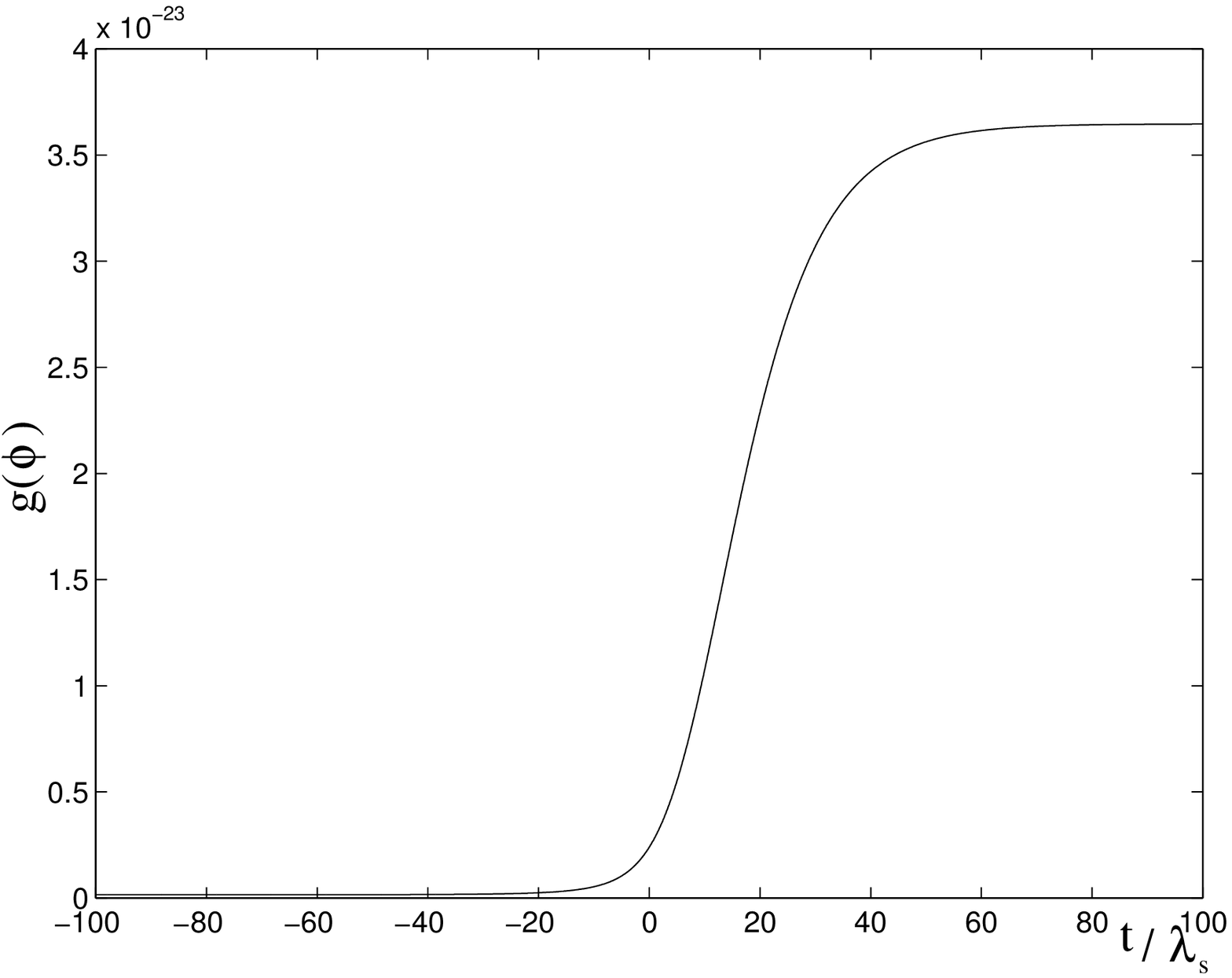}}&
      \hbox{\epsfxsize = 7.5 cm  \epsffile{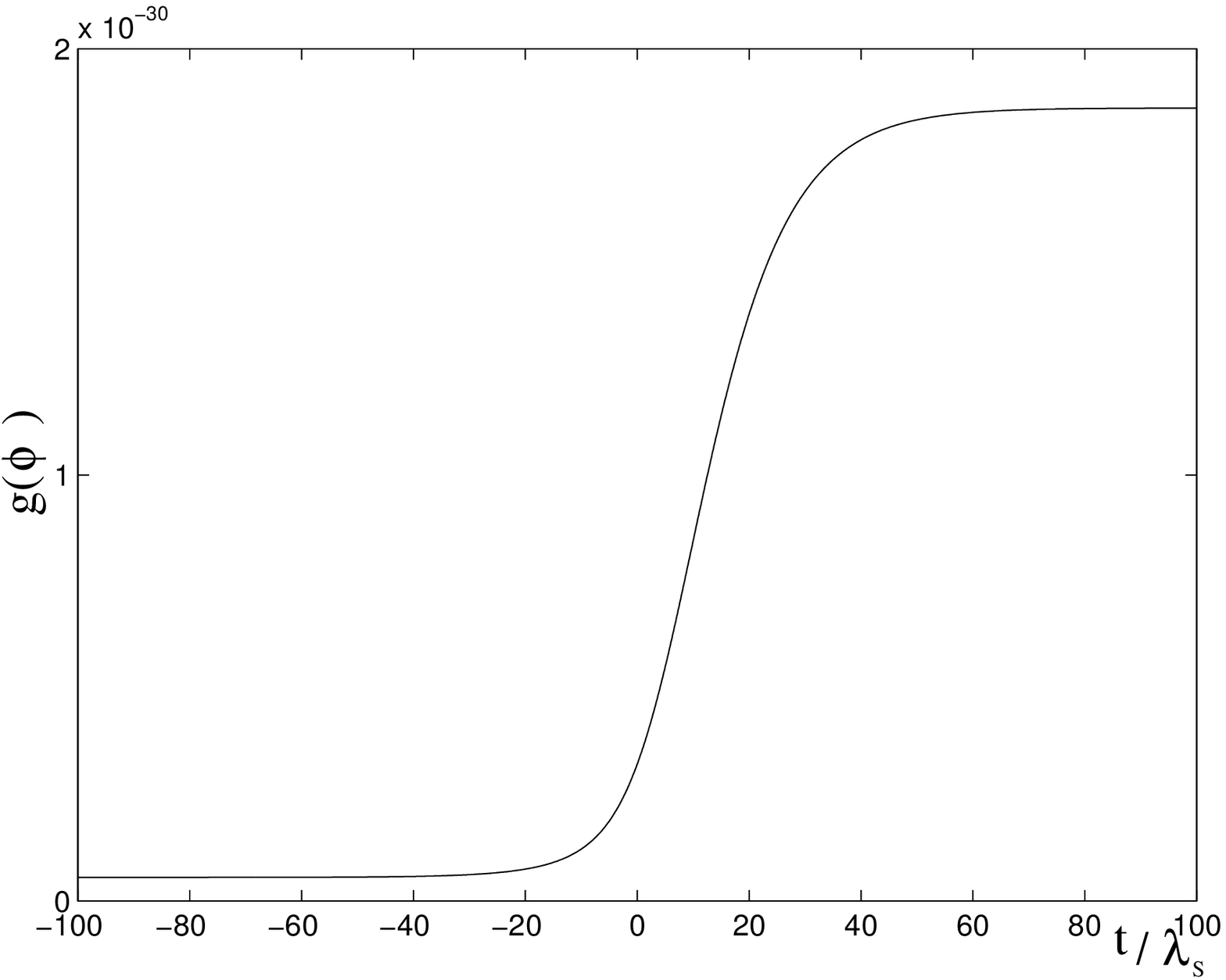}}\\
      \hline
\end{tabular}
\end{center}
\caption{We report the evolution of the dilaton coupling in the
solutions given by Eq. (\ref{thesolution1}) (left) and by Eq. 
(\ref{thesolution2}) (right). We took $c_{5} = 10~\sqrt{3}~\pi$ (left)
and $c_{5} =10~\sqrt{5}~\pi$ (right). Provided the
dilaton coupling starts its evolution deep in the perturbative regime,
$g(\phi)\ll 1$ for every time.}
\label{couplingfig}
\end{figure}

From
Eq. (\ref{inv}) and from the various forms of $f(x,t)$ which can be
deduced from Appendix B, we notice that ${\cal I}$ does not have any
pole for any finite value od $x$ and $t$. Moreover, we point out  that for
large $x$ and $t$ the curvature invariants are suppressed. The
magnitude of the suppression depends upon $a$ and $b$ which are
reported in  Table 1 but can also be read-off directly from
Eqs. (\ref{riemann})--(\ref{scalar}). 
The only nasty  thing which could then occur is that for some (large)
value of $x$ and/or $t$ some of the invariants explode because
$f(x,t)$ grows much faster than the suppression factor, parameterised,
for each invariant, by $a$ and $b$. 
Keeping $x=x_c$ (where $x_{c}$ is some finite  value of $x$) 
 and leaving $t$ to run we find that $f(x_{c},t)\sim [\cosh{\mu t}]^4$
for large $t$. Vice-versa keeping $t$ frozen at $t_{c}$ (where $t_{c}$
is some finite value of $t$) we find that (in the worse case)
$f(x,t_{c}) \sim [\cosh{\frac{\mu x}{2}}]^{4}$. Finally, by varying
$x$ and $t$ simultaneously we see that $f(x,t)$ can grow, at most
like $[\cosh{\frac{\mu x}{2}}]^{2}~ [\cosh{\mu t}]^4$ or like  
 $[\cosh{\frac{\mu x}{2}}]^{4}~ [\cosh{\mu t}]^2$. Therefore we see
that the growth of $f(x,t)$ is always suppressed by the high
(negative) powers (i.e. $a$ and $b$)
 of the hyperbolic cosinus appearing in Eq. (\ref{inv}).
\begin{table}
\begin{center}
\begin{tabular}{|c|c|c|c|c|}
\hline
$c_{4} = 2~\sqrt{3}~\mu$&
$R_{\mu\nu\alpha\beta}R^{\mu\nu\alpha\beta}$&
$C_{\mu\nu\alpha\beta}C^{\mu\nu\alpha\beta}$&
$R_{\alpha\beta}R^{\alpha\beta}$&
$R^2$\\
\hline
$a$ & $16$ & $16$ & $16$ & $16$\\
\hline
$b$ & $84$ & $84$ & $82$ & $80$\\
\hline
\end{tabular}
\end{center}
\nonumber
\begin{center}
\begin{tabular}{|c|c|c|c|c|}
\hline
$c_{4} = ~~\sqrt{5}~\mu $&
$R_{\mu\nu\alpha\beta}R^{\mu\nu\alpha\beta}$&
$C_{\mu\nu\alpha\beta}C^{\mu\nu\alpha\beta}$&
$R_{\alpha\beta}R^{\alpha\beta}$&
$R^2$ \\
\hline
$a$ & $14$ & $14$ & $14$ & $14$\\
\hline
$b$ & $52$ & $52$& $50$ & $48$\\
\hline
\end{tabular}
\end{center}
\label{ab}
\caption{We summarize the exponents of the suppression factor appearing
in Eq. (\ref{inv}). The upper part of the table refers to the case of
Eq. (\ref{thesolution1}), whereas the lower part of the table refers
to the case of Eq. (\ref{thesolution2}). The explicit analytical
results leading to this table can be found in the Appendix B, by
setting, respectively, $\a\equiv c_{4}/\mu = 2\sqrt{3}$ and
$\a=c_{4}/\mu \sqrt{5}$ in Eqs. (\ref{riemann}-(\ref{scalar}).}
\end{table}
We want now to make few more physical comments concerning the validity
of the present solutions.
The only dimension-full parameter appearing in the solution
(\ref{thesolution1}) and (\ref{thesolution2}) is $\mu = 1/L $ where
$L$ is, for the moment, a length scale which completely specifies each
solution which describes the ``spreading''  of the curvature
invariants around the origin (i.e. $x=0$, $t=0$). 
The only fundamental scale appearing in Eq. (\ref{action}) 
is the string scale $\lambda_{s}$. 
Correspondingly there is  also a typical
energy connected with $\lambda_{s}$ which is the string energy, namely
$E_{s}= 1/\lambda_{s}$. The effective action (\ref{action}) which was
the starting point of our considerations holds for energies $E\ll
E_{s}$. In fact for $E\sim E_{s}$ higher order corrections in the
string tension $\alpha'= \lambda_{s}^{-2}$ should be included.
In order to be consistent with our approach we have to assume that the
typical scale $L$ has to be large compared to the string scale and
this means that $L> \lambda_{s}$. In the opposite case
(i.e. $L<\lambda_{s}$), $\lambda_{s}$ would not turn out as the
minimal length scale. Therefore if $L >\lambda_{s}$, we will also have
${\cal I}~<~ \lambda_{s}^{-4}$.
In  Fig. \ref{rsquare} the scalar curvature (squared) is reported
 for the two solutions given
in Eqs. (\ref{thesolution1}) and (\ref{thesolution2}) when  $L=
10~\lambda_{s}$. 
The corresponding  analytical results are  reported 
in Eq. (\ref{scalar}) choosing, respectively $\a = 2 \sqrt{3}$ and
$\a = \sqrt{5}$. From these analytical
expressions  we can clearly see that by tuning $L= 1/\mu$ for length
scales larger than $\lambda_{s}$,  $R^2$ decreases (in string units) and
its spreading around the origin increases.
We can also  look at the behaviour of
the time derivative of the dilaton which is illustrated ( Fig. 
\ref{dilatonfig}) and of the related coupling constant 
(Fig. \ref{couplingfig}).

The solutions derived in Section 3 and the examples reported in the
present Section can be
described either in the String frame or in the Einstein frame
\cite{MC}.
In the String frame the fundamental
parameter of the theory is the String length $\lambda_{s}$. Notice
that in this frame the String scale is truly a {\em constant}, whereas
the Planck scale evolves according to a relation which can be
parameterised (at low energy and low coupling) as 
$\lambda_{P}(t) = e^{\frac{\phi}{2}} \lambda_{s}$.
The Einstein frame is defined by the conformal transformation which
diagonalizes the gravitational action by decoupling the dilaton from
the Einstein-Hilbert Lagrangian.In the Einstein frame 
$\lambda_{P} \sim 10^{-33} {\rm cm}$ is {\em
constant} whereas the String scale evolves according to 
$\lambda_{s}(t) = e^{-\frac{\Phi}{2}}\lambda_{P}$
(where $\Phi$ is the Einstein frame dilaton).
The String frame and Einstein frame dilatons are  simply related
and, moreover, in four dimensions they are indeed equal, whereas the
String frame ($g_{\mu\nu}$) and Einstein frame ($G_{\mu\nu}$) metrics
are connected by a conformal rescaling involving the dilaton coupling
($g_{\mu\nu} = e^{\Phi} G_{\mu\nu}$).

We transformed our solutions in the Einstein frame and we investigated
the emerging picture which can be summarized by saying that the
curvature invariants are also regular and everywhere defined.
At this point, we would be
tempted to call the solutions we just discussed
``singularity-free''. In order to name a generic solution of the low energy
String effective action ``singularity-free'' we should discuss the
geodesic completeness of the metric both in the Einstein and in the String
frame pictures. From the point of view of the interpretation of our solutions,
moreover, the  geodesic completeness in the String frame seems even
more crucial, since strings moving in curved space follow geodesics in
the String frame but not in the Einstein frame \cite{sanchez}. 
To show that the geodesics do not hit a boundary in both frames is
anyway a quite long computational task. We prefer to postpone it to
further investigations mainly because the calculations reported here
are already quite heavy. We do not have, at the moment, any argument
either favouring or disfavouring the geodesic completeness and we
believe that this can only come out from a precise calculation.

There is a second very important issue. The Einstein frame picture 
of our solutions involves an energy momentum tensor which {\em does
not} violate any of the energy conditions which are part of the
hypotheses of the powerful Hawking-Penrose (HP) theorems \cite{haw,wald}.
 Therefore the
question which naturally arises is which other hypothesis of the
HP theorems is violated. There are indeed examples of
truly non-singular (inhomogeneous) solutions in general relativity
which have an energy-momentum tensor of a perfect fluid \cite{sing}. 
In that context it was rigourously proven that the so-called  initial
or boundary condition of the HP theorems was the one failing to
hold. This would mean, in our context, that we have to fine-tune 
(as we had!) the
initial conditions quite heavily in order not to bump into a
singularity. This point is indeed
closely connected with the behaviour of the geodesics and with their 
completeness and therefore it certainly requires further studies.
The same considerations should be performed in the String frame
which is perhaps the most useful for the physical intuition (as it was
noticed in the case of the pre-big-bang scenario \cite{ven1}). We are
not aware, at the moment, of any general singularity theorem in this
case and therefore we are not in condition of making any kind of
general statement going beyond the solutions we just presented.

\renewcommand{\theequation}{5.\arabic{equation}}
\setcounter{equation}{0}
\section{Parity invariant solutions with growing coupling} 

The two physical examples presented in the previous Section share a
quite amusing physical property, namely the invariance of the
 line element corresponding to each one of the solutions
given in Eqs. (\ref{thesolution1}) and (\ref{thesolution2}) under
discrete parity transformations. In fact
by computing, from Eq. (\ref{line}), the line element associated to the
examples reported in Eq. (\ref{thesolution1}) and (\ref{thesolution2})
we find that it is invariant under the transformation $x\rightarrow -
x$. Surprisingly enough, we will show that  parity invariant
metrics lead to regular curvature invariants. This will allow us to
define a  class of regular, inhomogeneous models with growing
coupling.

We start by noting that, from the structure of the metric written in
Eqs. (\ref{thesolution}) the diagonal elements of the metric tensor
given in Eq. (\ref{metric}) are even under parity transformations
provided the spatial ($x$) dependence appears in functions which are
themselves even under parity. Given the form of the metric
(\ref{thesolution}) we see that, in order to preserve parity, the
spatial dependence has to occur either through powers of the
hyperbolic cosinus (which is parity even) or through even powers of
the hyperbolic sinus. We then impose
the following two extra  conditions to the parameters
$c_1,~c_2,~,c_3,~c_4$ specifying the solutions (\ref{thesolution}):
\begin{equation}
c^2_2 + c_2 c_3 - \mu c_1 + \mu^2 =0,~~~~~c_2 + c_3 = \mu
\label{parity}
\end{equation}
In Eqs. (\ref{thesolution}) the first condition of Eq. (\ref{parity}) 
sets to zero the power of the hyperbolic sinus
appearing in $A(x,t)$ whereas the second condition sets to one the
power of the hyperbolic cosinus appearing in $C(x,t)$. Notice that
with this choices the components of the metric are all even under
$x\rightarrow - x$.
Therefore, the full system of conditions obeyed by the solution given
in Eq. (\ref{thesolution}) is 
\begin{eqnarray}
&&\biggl(\frac{c_{3}}{\mu}\biggr)^2 = 3 +
\biggl(\frac{c_2}{\mu}\biggr)^2 - 2 \biggl(\frac{c_1}{\mu}\biggr),
\label{a}\\
&&\biggl(\frac{c_{4}}{\mu}\biggr)^2 = 
\biggl(\frac{c_2}{\mu}\biggr)^2 - 2 \biggl(\frac{c_1}{\mu}\biggr) -1,
\label{b}\\
&&\biggl(\frac{c_{2}}{\mu}\biggr)^2 = \biggl(\frac{c_1}{\mu}\biggr) - 
\biggl(\frac{c_2}{\mu}\biggr) \biggl(\frac{c_3}{\mu}\biggr),
\label{c}\\
&&\biggl(\frac{c_3}{\mu}\biggr) + \biggl(\frac{c_2}{\mu}\biggr) =1.
\label{d}
\end{eqnarray}
Now calling, for simplicity $\alpha = c_4/\mu$, $\gamma = c_2/\mu$,
$\delta= c_1/\mu$ and and eliminating $c_{3}/\mu$ through Eq. (\ref{a})
from the previous equations we obtain the following system:
\begin{eqnarray}
&&\gamma +1 = \delta, 
\label{a1}\\
&&\alpha^2 = \gamma^2 -1 - 2 \delta,
\label{b1}\\
&&2\gamma + 2 = 2 \delta.
\label{c1}
\end{eqnarray}
Eqs. (\ref{a1}) and (\ref{c1}) are clearly not independent
and, therefore, the solution given in Eq. (\ref{thesolution}) can be
expressed only in terms of one (physical) parameter which we choose to
be $\alpha$. This turns out to be a physically interesting
parametrization since $\mu^2 \alpha^2 = \dot{\phi}^2(0)$. In this sense 
$|\alpha|$ estimates the (maximal) energy density of the dilaton
background in string units.
 The sign of $\alpha$ automatically selects solutions
either with growing coupling or with decreasing coupling.
Thus, using Eqs. (\ref{d}), (\ref{a1}) and (\ref{b1}), we have that
the metric of Eq. (\ref{thesolution}) can be re-written as
\begin{eqnarray}
A(x,t) &=& e^{\pm\alpha~ gd(\mu t)}~ \Biggl[\cosh{\mu
t}\Biggr]^{2 \pm \sqrt{\alpha^2 + 4}}
\Biggl[\cosh{\left(\frac{\mu x}{2}\right)}
\Biggr]^{ 2\sqrt{\alpha^2 + 4}[\sqrt{\alpha^2 + 4} \pm 1] },
\nonumber\\
B(x,t) &=&  e^{\pm\alpha~gd(\mu t)}~\cosh{\mu t}~ \sinh{\mu x},
\nonumber\\
C(x,t) &=&\Biggl[\cosh{\mu t}\Biggr]^{1 \pm \sqrt{\alpha^2 + 4}}~ 
\Biggl[\sinh{\left(\frac{\mu x}{2}\right)}\Biggr]
\Biggl[\cosh{\left(\frac{\mu x}{2}\right)}\Biggr]^{1 \pm
2\sqrt{\alpha^2 + 4}},
\nonumber\\
\phi(t) &=& \pm\alpha ~gd(\mu t) - \beta.
\label{theclass}
\end{eqnarray}
(the plus and minus signs correspond, respectively, to solutions with
growing and decreasing dilaton coupling). In Appendix B we computed
the curvature invariants in the case of growing coupling
solutions. We indeed found that the requirement of parity invariance of
the solutions seem to be winning since the curvature invariants are
regular. Notice that the curvature invariants of the 
two  previous examples (reported in Section 4)
can be derived from the general result of this Section by setting,
respectively, $\a = 2 \sqrt{3}$ and $\a = \sqrt{5}$. 
Also in the case of the class of solutions discussed in the present
Section the curvature invariants vanish asymptotically as can be
argued from the results reported in Appendix B. 
The curvature invariants can be also computed in the case of
decreasing dilaton solution. We did this exercice and we found, again
regular solutions. From the more mathematical results reported in this
Section (and from the two  examples of Section 4 illustrating
the physical properties of our solutions) we see that something quite
unexpected happens, namely the fact that the regularity of the
curvature invariants can already be achieved at tree level (without
any curvature correction). Moreover, as shown in Section 4 for two
particular cases, we can set initial conditions in such a way that the
coupling constant is always much smaller than one. We notice that
these results are not in contrast with previous discussions
\cite{brusven} suggesting that the only way of regularizing the
curvature invariants was to invoke higher curvature corrections. In
fact those studies were always postulating a completely homogeneous
(and often isotropic) four dimensional geometry. The physical picture
emerging from our detailed analytical calculations is that the
inhomogeneities are quite an essential ingredient in order to
regularize the curvature invariants. On a purely physical ground,
moreover, it does not seem unreasonable to have strong inhomogeneities
at very small scales. 
As we noted in the introduction
our results confirm and extend previous ideas concerning possible
avoidance of the singularities in completely inhomogenous string
cosmological models \cite{GMV}. In agreement with \cite{GMV} we showed
that regular solutions are possible at tree-level. At the same time we
showed that this conclusion holds also in the presence of a growing
coupling solution and in the absence of any extra field.  We believe
that our new results should be taken as a serious motivation in order to
perform a fully consistent numerical study of inhomogeneous string
cosmological models. These studies might extend (and perhaps change)
some of the conclusions drawn in the completely homogeneous case 
\cite{brusven}. In particular, numerical studies might clarify if it
is at all possible to find inhomogeneous examples evolving at late
times towards a completely homogeneous state. In fact a quite
unphysical feature shared by our models and by the ones of \cite{GMV}
is their eternal inhomogeneity. It is certainly tue \cite{kra} that the
homogeneous limit of a inhomogeneous model is already quite delicate
in general relativity. In spite of the possible subtleties we can say
that a good measure of how inhomogeneous and anisotropic is a solution
at late times can be the ratio of the Weyl invariant 
($C^{\mu\nu\alpha\beta}C_{\mu\nu\alpha\beta}$) over the Riemann
invariant ($R^{\mu\nu\alpha\beta}R_{\mu\nu\alpha\beta}$). In our case,
 from the results of Appendix B, this quantity goes to a constant at
for $t\rightarrow +\infty$. A similar behaviour can be found in \cite{GMV}.
Numerical studies could also give us some clue concerning the role
played by the compactification radii which we took frozen but which
should be included in the game of inhomogeneous models. Up to now this 
possibility was never considered.

\renewcommand{\theequation}{6.\arabic{equation}}
\setcounter{equation}{0}
\section{Concluding remarks} 

In this paper we gave some  examples of regular dilaton
driven solutions of the low energy beta functions without including
any dilaton potential, any higher $\alpha'$ (curvature) 
corrections, any higher
genus corrections, any antisymmetric tensor field any dilaton
potential. We took
the six compactification radii to be constant and we looked at some
special class of inhomogeneous metric. 
We showed that if the metric is sufficiently inhomogeneous the
curvature invariants are bounded for every $x$ and $t$.
The only definite conclusion that we feel like stating
is that regular (but inhomogeneous)
solutions of the evolution equations of the dilaton and of the
geometry are more common (at tree-level) than in the completely
homogenous case where no-go theorems probably apply. 
Since in our examples the coupling is always very
small and the curvature invariants always smaller than one (in string
units) the tree-level treatment seems, a posteriori, not
meaningless. The
curvature invariants (including the Weyl invariants) vanish,
asymptotically for large times. Moreover the metric is invariant under
parity transformations.

We leave to  future studies  many unanswered questions.
Can we promote the regular solutions we spotted to true singularity
free solutions? If this is the case are the geodesics complete in the
Einstein frame? Which hypotheses of the singularity theorems
turn out to be violated? Can we formulate the singularity theorems (in
a consistent way) directly in the String frame or we have always to go to the
Einstein frame? Should we consider regular only those solutions which
are bounded in both frames?

\section*{Acknowledgments}
The author would like to thank M. Gasperini and 
 G. Veneziano for a careful reading of
the draft  and for very useful comments. The author wishes also to thank 
former discussions with H. Soleng.

\begin{appendix}

\section*{APPENDIX}

\renewcommand{\theequation}{A.\arabic{equation}}
\setcounter{equation}{0}
\section{Christoffel symbols and Ricci tensors} 

In this Appendix we will give the general formulas for the Christoffel
symbols and Ricci tensors computed in the case of the  diagonal metric
given in Eq. (\ref{metric}) 
with two commuting space-like killing vectors which are hypersurface
orthogonal. 

The Christoffel symbols can be easily derived from the metric
(\ref{metric}) and they are reported in  Table {\ref{tab1}},
whereas the  Ricci tensors and the curvature scalar 
can be obtained, after some algebra, once the Christoffel symbols are
known. Our conventions for the Ricci tensor will be 
$R_{\mu\nu}= R_{\alpha\mu\nu}^{~~~~\alpha}$ (where
$R_{\mu\nu\alpha}^{~~~~~\beta}=
\partial_{\mu}\Gamma_{\nu\alpha}^{~~~\beta}-...$ is the Riemann tensor).
The $(xx)$, $(yy)$, $(zz)$ and $(00)$ components of the Ricci tensors 
are, respectively
\begin{eqnarray}
&&R_{x}^{x} = - \frac{1}{2~A}\Biggl\{ - 
(\frac{\dot{A}}{A})^2 + \frac{\dot{A}}{A}\frac{\dot{B}}{B} +
 \frac{\ddot{A}}{A} + (\frac{A'}{A})^2 + \frac{A'}{A}\frac{B'}{B} 
+(\frac{B'}{B})^2 - (\frac{C'}{C})^2 - \frac{A''}{A} - 2 \frac{B''}{B}\Biggr\},
\label{rxx}\\
&&R_{y}^{y} = - \frac{1}{2~A}\Biggl\{ \frac{\dot{B}}{B}\frac{\dot{C}}{C} - 
(\frac{\dot{C}}{C})^2 +
\frac{\ddot{B}}{B} + \frac{\ddot{C}}{C} - \frac{B'}{B} \frac{C'}{C} +
(\frac{C'}{C})^2 - \frac{B''}{B}
- \frac{C''}{C}\Biggr\},
\label{ryy}\\
&&R_{z}^{z} = -\frac{1}{2~A} \Biggl\{ (\frac{\dot{C}}{C})^2 
- \frac{\dot{B}}{B}\frac{\dot{C}}{C}  
 + \frac{\ddot{B}}{B} - \frac{\ddot{C}}{C} 
+ \frac{B'}{B}\frac{C'}{C} - (\frac{C'}{C})^2 - \frac{ B''}{B} 
+ \frac{C''}{C}\Biggr\},
\label{rzz}\\
&&R_{0}^{0} = \frac{1}{2~A} \Biggl\{(\frac{\dot{A}}{A})^2 +
\frac{\dot{A}}{A}\frac{\dot{B}}{B} + (\frac{\dot{B}}{B})^2 - (
\frac{\dot{C}}{C})^2 
-\frac{\ddot{A}}{A} -2 \frac{\ddot{B}}{B} - (\frac{A'}{A})^2 +
\frac{A'}{A}\frac{B'}{B} + \frac{A''}{A}\Biggr\},
\label{r00}
\end{eqnarray}
The $(0x)$ component and the curvature scalar will be instead:
\begin{equation}
R_{x}^{0} = \frac{1}{2~A} \Biggl\{ \frac{\dot{B}}{B}\frac{A'}{A} +
\frac{\dot{A}}{A}\frac{B'}{B} +  \frac{B'}{B} \frac{\dot{B}}{B} - 
\frac{\dot{C}}{C}\frac{ C'}{C} - 2 \frac{{\dot{B}}'}{B} \Biggr\},
\label{r0x}
\end{equation}
\begin{equation}
R  \equiv R_{\alpha}^{\alpha} = \frac{1}{2~A} \Biggl\{2 (\frac{\dot{A}}{A})^2
+(\frac{\dot{B}}{B})^2 - (\frac{\dot{C}}{C})^2 
- 2\frac{\ddot{A}}{A} - 4 \frac{\ddot{B}}{B}
 - 2(\frac{A'}{A})^2 -(\frac{B'}{B})^2 +
(\frac{C'}{C})^2 + 2 \frac{A''}{A} + 4 \frac{B''}{B}\Biggr\}.
\label{r}
\end{equation}
\begin{table}
\begin{center}
\begin{tabular}{|c|c|}
\hline
 $\Gamma_{x x}^{x} = \frac{1}{2}\frac{\partial \log{A}}{\partial x}$ &
 $\Gamma_{yy}^{x}  =
 -\frac{C~B}{2~A}\frac{\partial \log{(B~C)}}{\partial x}  $ \\
\hline
 $ \Gamma_{zz}^{x} = - \frac{B}{2~ A ~C} \frac{\partial\log{(B~C)}}{\partial x}
 $ & $\Gamma_{x 0}^{x} = \frac{1}{2}\frac{\partial\log{A}}{\partial~t} $\\
\hline
 $\Gamma_{00}^{x} =\frac{1}{2}\frac{\partial \log{A}}{\partial x}$ & 
$\Gamma_{xy}^{y}=
 \frac{1}{2}\frac{\log{(B~C)}}{\partial x}$ \\
\hline
 $\Gamma_{y0}^{y} = \frac{1}{2}\frac{\log{(B~C)}}{\partial t} $ &
$\Gamma_{xz}^{z}=
 \frac{1}{2}\frac{\partial\log{(B/C)}}{\partial x} $ \\
\hline
 $  \Gamma_{z0}^{z}=\frac{1}{2}\frac{\partial\log{(B/C)}}{\partial t}  $ 
& $\Gamma_{xx}^{0}=\frac{1}{2} \frac{\partial\log{A}}{\partial t}$ \\
\hline
$ \Gamma_{yy}^{0}= \frac{C~B}{2~A}\frac{\partial\log{(C~B)}}{\partial t}$ &
$ \Gamma_{zz}^{0}=\frac{B}{2~A~C} \frac{\partial\log{(B/C)}}{\partial t}$\\
\hline
$\Gamma_{x0}^{0} = \frac{1}{2} \frac{\partial\log{A}}{\partial x}$& $
\Gamma_{00}^{0} 
=\frac{1}{2}\frac{\partial\log{A}}{\partial t}$\\
\hline
\end{tabular}
\end{center}
\caption{We report the Christoffel symbols computed from the metric
given in Eq. (\ref{metric}).}
\label{tab1}
\end{table} 

\renewcommand{\theequation}{B.\arabic{equation}}
\setcounter{equation}{0}
\section{Curvature invariants for growing dilaton solutions} 

From the (parity invariant) class of solutions given in
Eq. (\ref{theclass}) we can compute the curvature invariants, namely
curvature scalar together with 
the squares of the Riemann, Weyl and Ricci tensors. The result is the 
following
\begin{eqnarray}
R_{\mu\nu\alpha\beta}R^{\mu\nu\alpha\beta} &=& \frac{\mu^4}{32} 
~e^{- 2\alpha gd(\mu t)} [\cosh{\mu t}]^{-8 - 2 \sqrt{\alpha^2 +4}}
\Biggl[\cosh{\frac{\mu x}{2}}\Biggr]^{-4 - 4\sqrt{\alpha^2 + 4}(1 +
\sqrt{\alpha^2 +1})}
\Biggl\{ s_1(\alpha)  
\nonumber\\
&+& s_2(\alpha) \cosh{ 2 \mu t} +s_{3}(\alpha)
\cosh{4 \mu t} +  s_4(\a) \cosh{\mu x} + s_5(\alpha) \cosh{2 \mu x} 
\nonumber\\
&+& s_6(\a) \cosh{2 \mu t} \cosh{2\mu x} + s_7(\a) \cosh{2 \mu t} \cosh{\mu
x}  
\nonumber\\
&+& s_8(\a) \cosh{4 \mu t} \cosh{\mu x}+ s_9(\a) \sinh{\mu t}
+ s_{10}(\a)  \sinh{ \mu t}\cosh{2  \mu x}  
\nonumber\\
&+& s_{11}(\a) \sinh{\mu t} \cosh{ \mu x} \Biggr\},
\label{riemann}\\
C_{\mu\nu\alpha\beta}C^{\mu\nu\alpha\beta} &=& \frac{\mu^4}{96} ~e^{- 2\a
gd(\mu t)}~[\cosh{\mu t}]^{-8 - 2 \sqrt{\a^2 + 4}}~\Biggl[\cosh{\frac{\mu
x}{2}}\Biggr]^{ -4 - 4 \sqrt{\a^2 + 4}(1 + \sqrt{\a^2 + 4})} 
\Biggl\{w_{1} (\a) 
\nonumber\\
&+& w_{2}(\a) \cosh{4 \mu t} +
w_{4}(\a) \cosh{\mu x} + w_{5}(\a) \cosh{2 \mu x}
\nonumber\\
&+& w_{6}(\a) \cosh{2\mu t} \cosh{2\mu x}
+ w_{7}(\a) \cosh{2\mu t} \cosh{\mu x}
\nonumber\\
&+& w_{8}(\a) \cosh{4 \mu t} \cosh{\mu x}\Biggr\},
\label{weyl}\\
R_{\alpha\beta}R^{\alpha\beta} &=&\frac{\a^2 \mu^4}{16}~ e^{-2\a gd(\mu t)}
~[\cosh{\mu t}]^{-8-2\sqrt{\a^2 +4}} ~\Biggl[\cosh{\frac{\mu x}{2}}
\Biggr]^{-2 - 4 \sqrt{\a^2 + 4} (1 + \sqrt{\a^2 + 4})}
~\Biggl\{ r_1(\a)  
\nonumber\\
&+&r_{2} (\a) \cosh{2\mu t} + r_{3}(\a) \cosh{\mu x} + r_{4}(\a)
\cosh{2\mu t} \cosh{\mu x}
\nonumber\\
&+& r_5(\a) \sinh{\mu t} + r_{6}(\a) \sinh{\mu t}
\cosh{\mu x} \Biggr\},
\label{ricci}\\
R^2 &=& \a^4 ~\mu^4 ~e^{- 2\a gd(\mu t)} ~[\cosh{\mu t}]^{-8
-2\sqrt{\a^2 + 4}}
\Biggl[\cosh{\frac{\mu x}{2}}\Biggr]^{-4 \sqrt{\a^2 + 4} (1 +
\sqrt{\a^2 + 4})}.
\label{scalar}
\end{eqnarray}

The $\alpha$-dependent coefficients appearing in our expressions 
 are reported in Table
\ref{tab3}.
Notice that the two examples specifically discussed in Section 4 arise
by setting, respectively, $\a = 2\sqrt{3}$ and $\a =
\sqrt{5}$. Plugging these values in
Eqs. (\ref{riemann})--(\ref{scalar}) we get the corresponding
curvature invariants. We want to mention that we did the calculations
 of the curvature invariants also in the Einstein frame after having
 transformed the solutions according to the general conventions
 (discussed in Section 4) which relate the two frames. 
 The curvature invariants turned out to be regular also in the
 Einstein frame.

\begin{table}
\begin{center}
\begin{tabular}{|c|c|}
\hline
\hline
$s_1(\a)$ &
$ 1152 + 416 \a^2 + 824 \a^4 + 6 \a^6  + \sqrt{ \a^2 + 4}( 504 + 98
 \a^2 + 14 \a^4) $ \\
\hline
$s_2(\a)$ &
$ 384 + 536 \a^2 + 124 \a^4 + 8 \a^6 + \sqrt{\a^2 + 4} ( 96 + 164 \a^2
+ 20 \a^4)$\\
\hline
$s_3(\a)$ & $ 384 + 216 \a^2 + 38 \a^4 + 2 \a^6  + \sqrt{\a^2 + 4} (
168 + 66\a^2 + 6\a^4)$\\
\hline
$s_4(\a)$ & $ 48 - 108 \a^2 - 78\a^4 - 6 \a^6  - \sqrt{\a^2 +4}( 24 +
190\a^2 + 18\a^4)$\\
\hline
$s_{5}(\a)$ &$ 192 + 104 \a^2 + 16\a^4 + \sqrt{\a^2 + 4}(192 + 36 \a^2
+ 4 \a^4)$\\
\hline
$s_{6}(\a)$ & $ 72 \a^2 + 12\a^4 + \sqrt{\a^2 + 4}( 36 \a^2 +
4\a^4)$\\
\hline
$s_{7}(\a)$ & $ -1728 - 688 \a^2 -120 \a^4 - 8\a^6 - ( 864 + 232 \a^2
+ 24 \a^4 ) \sqrt{\a^2 + 4})$\\
\hline
$s_{8}(\a) $ & $ -144 - 108 \a^2 - 18\a^4 - (72 + 42 \a^2 +
6\a^4)\sqrt{\a^2 + 4}$\\
\hline
$s_{9}(\a) $ & $ 96 \a^3 + 24\a^3 \sqrt{\a^2 + 4}$\\
\hline
$s_{10}(\a) $ & $ 32\a^3 + 8\a^3 \sqrt{\a^2 + 4}$\\
\hline
$s_{11}(\a) $ & $ 128 \a^3 + 32 \a^3 \sqrt{\a^2 + 4}$\\
\hline
\hline
$w_{1}(\a) $ & $ 3456 + 1488 \a^2 + 216 \a^4 + 12 \a^6 + \sqrt{\a^2 +
4} ( 1512 + 390 \a^2 + 30 \a^4)$\\
\hline
$w_{2}(\a) $ & $ 1152 + 1128 \a^2 + 282 \a^4 + 18 \a^6 + \sqrt{\a^2 +
4} ( 288 + 300 \a^2 + 48 \a^4 ) $\\
\hline
$w_{3}(\a) $ & $ 1152 + 648 \a^2 + 114 \a^4 + 6 \a^6 + \sqrt{\a^2 + 4}
(504 + 198 \a^2 + 18 \a^4)$\\
\hline
$w_4(\a) $ & $ -144 - 492 \a^2 - 266 \a^4 - 18 \a^6 - ( 72 + 378 \a^2 +
54 \a^4)\sqrt{ \a^2 + 4}$\\
\hline
$w_{5}(\a) $ & $ 1152 + 552 \a^2 + 94 \a^4 + 6 \a^6 + \sqrt{\a^2 + 4}
(576 + 204 \a^4 +  24 \a^4)$\\
\hline
$w_{6}(\a) $ & $ 216 \a^2 + 74 \a^4 + 6 \a^6 + \sqrt{\a^2 + 4} ( 108
\a^2 + 24 \a^4 ) $ \\
\hline
$w_{7}(\a) $ & $ - 5184 -1824 \a^2 -1128 \a^4 -24 \a^6 - \sqrt{\a^2 +
4} ( 2592 + 888 \a^2 + 72 \a^4)$\\
\hline
$w_{8}(\a) $ & $ -( 216 + 126 \a^2 + 18\a^4 ) \sqrt{\a^2 + 4}$ \\
\hline
\hline
$r_{1}(\a)$ & $ 22 \a^2 + 2\a^4 + 4\a^2 \sqrt{\a^2 + 4}$\\
\hline
$r_{2}(\a)$ & $ 80 + 22\a^2 + 2\a^4 + (32 + 4\a^2)\sqrt{\a^2 + 4}$\\
\hline
$r_{3}(\a)$ & $ -80 -14 \a^2 - 2\a^4 - (32 + 4\a^2) \sqrt{\a^2 + 4}$\\
\hline
$r_{4}(\a)$ & $ -14 \a^2 -2 \a^4 - 4\a^2 \sqrt{\a^2 +4}$\\
\hline
$r_{5}(\a)$ & $ 32 \a + 8\a \sqrt{\a^2 + 4}$\\
\hline
$r_{6}(\a)$ & $ 32 \a + 8\a \sqrt{\a^2 +4}$\\
\hline
\hline
\end{tabular}
\end{center}
\caption{We report the coefficients appearing in the expression of
the Riemann, Weyl  and Ricci  invariants given, respectively, in Eqs. 
(\ref{riemann}), (\ref{weyl}) and (\ref{ricci}).}
\label{tab3}
\end{table} 

\end{appendix}

\end{document}